# Extremely intrinsic chirality in two-dimensional planar waveguide grating induced by quasi-bound states in the continuum

Dandan Zhang[1], Tingting Liu[2], Linlin Lei[3], Weimin Deng[1], Tongbiao Wang[1], Qinghua Liao[1], Wenxing Liu[1,*], Shuyuan Xiao[2,†], and Tianbao Yu[1,‡]

[1]School of Physics and Materials Science, Nanchang University, Nanchang 330031, China
[2]Institute for Advanced Study, Nanchang University, Nanchang 330031, China
[3]Jiujiang vocational and technical college, Jiujiang 332007, China

*liuwenxing@ncu.edu.cn
†syxiao@ncu.edu.cn
‡ yutianbao@ncu.edu.cn

**Abstract:** The strong chiral light-matter interaction is crucial for various important fields such as chiral optics, quantum optics, and biomedical optics, driving a quest for the extreme intrinsic chirality assisted by ultrahigh quality ($Q$-) factor resonances. In this quest, we propose a straightforward method to achieve extreme intrinsic chirality in lossless planar structures by manipulating the quasi-BIC through in-plane perturbation. The temporal coupled-mode theory is employed to derive the conditions necessary for achieving maximal intrinsic chirality. The quasi-BIC should be excited within the transparent spectral range of the structure and couple with $x$- and $y$-polarized waves with the same intensity but a phase difference of $\pi/2$. For an illustration, a planar chiral dielectric dimeric waveguide grating is designed that strong interacts with left circularly polarized (LCP) light while decouples from right circularly polarized (RCP) light through in-plane symmetry engineering. Furthermore, by adjusting the magnitude of the in-plane asymmetry, we can independently manipulate the $Q$-factors of the chiral quasi-BIC while maintaining nearly unity circular dichroism. Our results provide a simple yet powerful paradigm for achieving extreme intrinsic chirality on an easily manufacturable platform, which may have potential applications in chiral emission, chiral sensing, and enantiomer separation.

## 1. Introduction

Chirality refers to the geometric property of an object that cannot be superimposed onto its mirror image by rotation or translation[1]. In the context of optics, chiral objects interact with left/right circularly polarized (LCP/RCP) light can induce chiroptical effects, including optical activity and circular dichroism. Optical devices with a strong chiroptical response are crucial in various applications, such as chiral molecule sensing[2], enantiomer selection[3], and quantum optics[4]. However, the chiral light-matter interactions in natural materials are typically very weak, making it challenging to detect the resulting optical response. In recent years, benefiting from the unprecedented flexibility for design, various chiral nanophotonic structures have been considered to generate strong chiral optical responses, such as the helices[5, 6], twisted cross structures[7, 8], and multi-layered structures[9, 10]. Nevertheless, due to absorption and scattering losses, these structures only support low $Q$-factor optical responses, which inevitably hampers their potential applications relying on strong chiral light-matter interactions.

Recently, the emergence of bound states in the continuum (BICs) has provided a viable alternative to enhance the strength of chiral light-matter interactions, owing to their unique ability to confine energy[11-14]. BICs are localized photonic eigenstates embedded in the radiation continuum, which can be considered as a resonance with zero linewidth or an infinite $Q$-factor due to the energy be perfectly confined within the system[15, 16]. Typically, the limited size of the structure, material losses, and other external perturbations can cause BICs to collapse into Fano resonance with finite but high $Q$-factors, which are referred to as quasi-BICs[17-23]. The quasi-BICs with significantly high $Q$-factors have been applied in areas such as low-threshold lasing[24-26], efficient nonlinear optical processes[27-29], and unidirectional emission[30-33]. Importantly, when a quasi-BIC acquires intrinsic chirality, the resulting chiral quasi-BIC can generate strong chiroptical effects accompanying with high $Q$-factors. For example, a bilayer twisted structure could efficiently control chirality due to the absence of all mirror symmetries[34]. The maximum intrinsic chirality can be achieved by introducing out-of-plane perturbations to control the coupling of quasi-BIC with circularly polarized waves, as exemplified by pairs of ellipse dimers with a small out-of-plane distance[35-37], tilted etch twisted dimers[31], and slanted trapezoid nanoholes in dielectric film[38]. However, the introduced out-of-plane perturbations resort to three-dimensional sophisticated geometries, which is challenging to conventional 2D photolithography manufacturing techniques.

In this work, we propose a straightforward method to achieve the extreme intrinsic chirality in planar structures through manipulating the quasi-BIC *via* in-plane perturbation. Based on the temporal coupled-mode theory (CMT), we derive the conditions necessary for achieving maximal intrinsic chirality: i) the quasi-BIC resonance should occur within the transparent spectral range of the structure, ii) the dissipation loss can be negligible, and iii) the quasi-BIC is to couple with $x$- and $y$-polarized waves of the same intensity but with a phase difference of $\pi/2$. As an example, we designed a planar waveguide grating comprising a dimeric grating and a waveguide layer that fulfill the above conditions. Through in-plane symmetry engineering, a high-$Q$ guided mode resonance corresponding to the quasi-BIC can be excited, which exhibits strong interaction with LCP light while decoupling from RCP light. The simulation results demonstrate 99.88% transmittance and reflectance for RCP and LCP waves, respectively. Moreover, by adjusting the magnitude of the in-plane asymmetry, we can independently manipulate the $Q$-factors of the chiral quasi-BIC while maintaining nearly unity circular dichroism (CD). These results have significant potential spin-selective devices and applications.

## 2. Theoretical analysis of Maximum CD

To precisely manipulate the quasi-BIC resonance in circular polarization base, we start by discussing the necessary condition for maximal chirality based on Jones matrix analysis. The transmission matrices between the circular and linear polarization bases are related as [39]:

$$\mathbf{T}_{\text{circ}} = \begin{pmatrix} t_{RR} & t_{RL} \\ t_{LR} & t_{LL} \end{pmatrix} = \frac{1}{2} \begin{pmatrix} (t_{xx}+t_{yy})+i(t_{xy}-t_{yx}) & (t_{xx}-t_{yy})-i(t_{xy}+t_{yx}) \\ (t_{xx}-t_{yy})+i(t_{xy}+t_{yx}) & (t_{xx}+t_{yy})-i(t_{xy}-t_{yx}) \end{pmatrix}, \quad (1)$$

where the first and second subscripts refer to the transmitted and incident waves, $R$ and $L$ denote the right-handed circularly polarized wave (RCP) and left-handed circularly polarized wave (LCP), $x$ and $y$ represent the two orthogonal linearly polarized waves. The transmission CD is defined as $CD = |t_{RR}|^2 + |t_{LR}|^2 - |t_{RL}|^2 - |t_{LL}|^2$. The maximal CD requires that the chiral structures completely transmit one helicity of circularly polarized incident light but completely reflect circularly polarized incident light with the opposite handedness. We can readily prove that maximal CD follows Jones matrix in the linear and circular polarization bases

$$\mathbf{T}_{lin} = \begin{pmatrix} t_{xx} & t_{xy} \\ t_{yx} & t_{yy} \end{pmatrix} = \frac{e^{i\alpha}}{2}\begin{pmatrix} 1 & i \\ i & -1 \end{pmatrix}, \mathbf{T}_{circ} = e^{i\alpha}\begin{pmatrix} 0 & 1 \\ 0 & 0 \end{pmatrix}, \quad (2)$$

$$\mathbf{T}_{lin} = \begin{pmatrix} t_{xx} & t_{xy} \\ t_{yx} & t_{yy} \end{pmatrix} = \frac{e^{i\alpha}}{2}\begin{pmatrix} 1 & -i \\ i & 1 \end{pmatrix}, \mathbf{T}_{circ} = e^{i\alpha}\begin{pmatrix} 1 & 0 \\ 0 & 0 \end{pmatrix}. \quad (3)$$

where $\alpha$ is an arbitrary phase shift through the chiral structure. For the case of Eq. (2), the CD originates from anisotropy-induced polarization conversion, which is usually referred to as 'false chirality' [38, 40-45]. In the case of Eq. (3), only one helicity of circular polarization incidence can be transmitted without polarization conversion, which is considered as 'true chirality' or 'intrinsic chirality' [38]. In this study, we only consider the intrinsic chirality.

To reveal the mechanism and formation of extreme chirality in chiral structure, we resort to phenomenological coupled mode theory (CMT) with two ports. We begin by analyzing the most general transmission-reflection problem within the S-matrix in the basis of linear polarizations

$$\mathbf{S} = \begin{pmatrix} r_{1xx} & t_{2xx} & r_{1xy} & t_{2xy} \\ t_{1xx} & r_{2xx} & t_{1xy} & r_{2xy} \\ r_{1yx} & t_{2yx} & r_{1yy} & t_{2yy} \\ t_{1yx} & r_{2yx} & t_{1yy} & r_{2yy} \end{pmatrix}, \quad (4)$$

where the $r_{mjk}$ and $t_{mjk}$ are the reflection and transmission amplitudes of $j$-component under incident $k$-component light from structure sides $m$. Here, we consider the chiral structure hosting a single resonance. According to the CMT, the dynamic equations for the amplitude A of the resonance mode can be written as [39, 46, 47]:

$$\frac{dA}{dt} = (-i\omega_0 - \gamma)A + \boldsymbol{\kappa}^T \mathbf{a}, \quad (5)$$

$$\mathbf{b} = \mathbf{Sa} = \mathbf{Ca} + \mathbf{MA}, \quad (6)$$

where $\omega_0$ and $\gamma$ are the center frequency and the decay rate of the resonance, respectively; $\boldsymbol{a} = \begin{bmatrix} a_{1x} & a_{2x} & a_{1y} & a_{2y} \end{bmatrix}^T$ and $\boldsymbol{b} = \begin{bmatrix} b_{1x} & b_{2x} & b_{1y} & b_{2y} \end{bmatrix}^T$ represent the amplitudes of incoming and outgoing waves, where the subscripts denote the side of the structure and polarizations; $\mathbf{C}$ is the scattering matrix for the background transmission and reflection through the structure; $\boldsymbol{\kappa} = \begin{bmatrix} m_{1x} & m_{2x} & m_{1y} & m_{2y} \end{bmatrix}^T$ describes the coupling between the resonance and the incoming waves. According to Lorentz reciprocity, the matrix $\boldsymbol{M}$ is related to $\boldsymbol{\kappa}$ and is given by $\boldsymbol{M} = \boldsymbol{\kappa}$. Naturally, the single resonance is coupled to all linearly polarized waves on both sides of the structure. The stationary solution of Eqs. (5) and (6) determines that the S-matrix from Eq. (4) as:

$$\mathbf{S} = \mathbf{C} - \frac{\mathbf{M}\boldsymbol{\kappa}^T}{i(\omega - \omega_0) - \gamma}. \quad (7)$$

From Eq. (7), we can obtain all the reflection and transmission amplitudes in Eq. (4) by the model parameters. To achieve the intrinsic chirality describing in Eq. (3), the structure usually needs to have $C_4$ rotational symmetry with respect to the $z$ axis[39]. However, the intrinsic chirality in $C_4$-symmetric structure is very weak, especially for a two-dimensional plane configuration[48, 49]. Interestingly, the intrinsic chirality in Eq. (3) could be satisfied when the structure hosts a resonance in a spectral range where the background is perfectly transparent. In the case, the scattering matrix of background is given by

$$\mathbf{C} = \begin{pmatrix} 0 & 1 & 0 & 0 \\ 1 & 0 & 0 & 0 \\ 0 & 0 & 0 & 1 \\ 0 & 0 & 1 & 0 \end{pmatrix}. \tag{8}$$

According to energy conservation and time-reversal symmetry, we have

$$\mathbf{C}\mathbf{\kappa}^* = -\mathbf{M}, \tag{9}$$

$$\mathbf{M}^*\mathbf{M} = 2\gamma. \tag{10}$$

From Eqs. (9) and (10), we can obtain the coupling parameters on different sides of the structure[50]: $m_{2x} = -m_{1x}^*$, $m_{2y} = -m_{1y}^*$, and $|m_{1x}|^2 + |m_{2x}|^2 + |m_{1y}|^2 + |m_{2y}|^2 = 2\gamma$. Due to Lorentz reciprocity, we only consider the transmission matrix from side 1 of the structure,

$$\mathbf{T}_{\text{lin1}} = \begin{pmatrix} t_{1xx} & t_{1xy} \\ t_{1yx} & t_{1yy} \end{pmatrix} = \begin{pmatrix} 1 + \dfrac{|m_{1x}|^2}{i(\omega-\omega_0)-\gamma} & \dfrac{|m_{1x}||m_{1y}|e^{i\Delta\Phi}}{i(\omega-\omega_0)-\gamma} \\ \dfrac{|m_{1x}||m_{1y}|e^{-i\Delta\Phi}}{i(\omega-\omega_0)-\gamma} & 1 + \dfrac{|m_{1y}|^2}{i(\omega-\omega_0)-\gamma} \end{pmatrix}, \tag{11}$$

where $\Delta\Phi$ is phase difference between the coupling parameters $m_{1x}$ and $m_{1y}$. From Eq. (11), we can obtain the necessary condition for the maximal intrinsic chirality in Eq. (3): $|m_{1x}| = |m_{1y}|$ and $\Delta\Phi = \pi/2$. The transmission matrix for circular polarization waves is then written as:

$$\mathbf{T}_{\text{circ}} = \begin{pmatrix} 1 & 0 \\ 0 & 1 + \dfrac{\gamma}{i(\omega-\omega_0)-\gamma} \end{pmatrix}, \tag{12}$$

where the maximal intrinsic chirality in the structure is achieved at the resonant frequency of $\omega = \omega_0$.

The above analysis concludes the necessary condition to realize the maximal intrinsic chirality in a structure: the resonance should occur in the spectral range of structure transparency, the dissipation in the structure material can be negligible, and the structure must couple with the *x*-polarized and *y*-polarized waves satisfying the given condition: $|m_{1x}| = |m_{1y}|$ and $\Delta\Phi = \pi/2$.

## 3. Design of planar waveguide grating with maximum chirality

To fulfill the above conditions of maximum intrinsic chirality, we designed a planar chiral waveguide grating as illustrated in Fig. 1. It is composed of periodic dimeric rectangular bars arranged on a waveguide layer. To achieve perfect transparency in the range of visible wavelengths, the structure is made entirely of TiO$_2$ (the permittivity is 5.76) and each unit cell is characterized by a period $\Lambda$=580 nm, width, length, and thickness of rectangular bars are *w*=80 nm, *l*=150 nm, $t_1$=220 nm, waveguide layer thickness $t_2$=140 nm. The presence of the waveguide layer breaks the structural symmetry in the *z*-axis direction, thereby providing a phase difference between $m_x$ and $m_y$. The distance between the centers of the two rectangular bars along the *x* direction is *Λ/2* at the original position. In our design, the in-plane symmetry of the structure is broken by moving one of the rectangular bars, and the distance of the move along the *x* and *y* directions is denoted as Δ*x* and Δ*y*, respectively. Such a planar structure can be manufactured using conventional photolithography techniques[51-53].

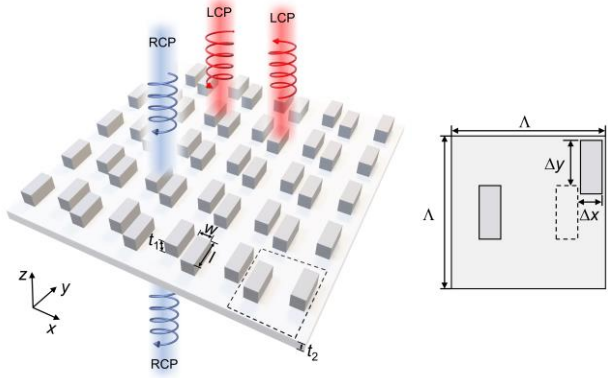

Fig. 1. The schematic configuration of the planar chiral waveguide grating composed of periodic dimeric rectangular bars arranged on a waveguide layer. The incident light is assumed to be a plane wave propagating along the *z* direction.

For the high *Q*-factor resonance in the spectral range of structure transparency, we exploit the engineered Brillouin zone folding-induced quasi-BIC. The first Brillouin zone (FBZ) of the structure is shown in Fig. 2(a). The dashed and solid boxes represent FBZs with and without perturbations, respectively. The blue shaded region indicates that the FBZ has been halved in the *x* direction due to the doubling of the period in the *x* direction. The momentum properties are simulated with commercial finite-element software COMSOL Multiphysics. In the simulations, the Floquet periodic boundary conditions are applied to a single unit cell, and the perfectly matched layers are used in the *z* direction. Here, we focus on a transverse-magnetic (TM)-like eigenmode in the waveguide layer. When Δ*x*=0 nm and Δ*y*=0 nm, the grating degenerates to a half-period grating along the *x* direction with $\Lambda' = \Lambda/2$, and its band corresponds to the red line in Fig. 2(b). It can be observed that the TM-like mode is below the light line (gray), indicating that the mode behaves as guided mode and cannot couple with the external incident light due to wavevector mismatch. While introducing a period perturbation along the *x* direction by moving one of the rectangular bars along the *x* or *y* direction, the grating period became Λ, and its band (blue) is folded from the band of the half-period structure (red). Then, the guided mode rises to the Γ point above the light line and couples with the external incident light, turning into resonance. Since the resonance is born from a small geometric perturbation, the coupling of the guided mode with the external incident light can be considered as a quasi-BIC resonance[13, 34, 54, 55].

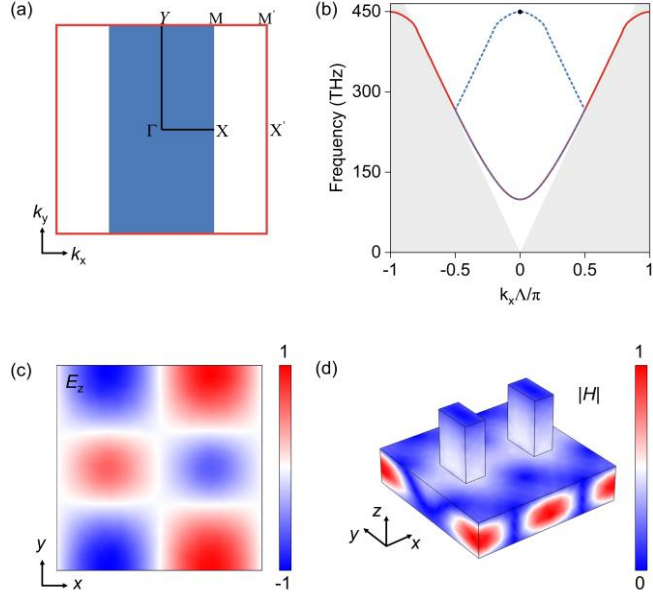

**Fig. 2.** (a) The first Brillouin zone of the structure. (b) The band structure of eigenmodes simulated with unit- (blue) and half- (red) period boundary, light cone is represented gray shaded area. (c) The eigen electric field distribution with $z$ component ($E_z$) at the $\Gamma$ point. (d) The eigen magnetic field distribution ($|H|$) at the $\Gamma$ point.

The corresponding electromagnetic field distributions at the $\Gamma$ point is shown in Fig. 2(c) and 2(d), respectively. As illustrated in Fig. 2(c), the $z$-component of the electric field exhibits an odd symmetry under a 180° rotation around the $z$-axis. The symmetries of the quasi-BIC can be tracked in the language of Group Theory by the irreducible representation $B_1$ mode of the point group $C_{2v}$. The selection rules show that the $B_1$ mode can radiate into free space in different polarization directions by introducing different perturbations[34, 54, 56]. When one of the rectangular bars moves along the $x$ direction, the breaking of the symmetry along the $x$ direction will introduce a leaky channel coupled to the continuum for $x$-polarized light. While for the rectangular bar moves along the $y$ direction, the structure lacks any in-plane mirror symmetries, other than in-plane inversion ($C_2$) symmetry, which will lead to the $y$-polarized light exhibits a leaky resonance. Here, the magnetic field ($|H|$) suggests that the guided mode is strongly localized in the waveguide layer as shown in Fig. 2(d). Therefore, the phase difference between the $m_x$ and $m_y$ will be slightly influenced by $\Delta x$ and $\Delta y$.

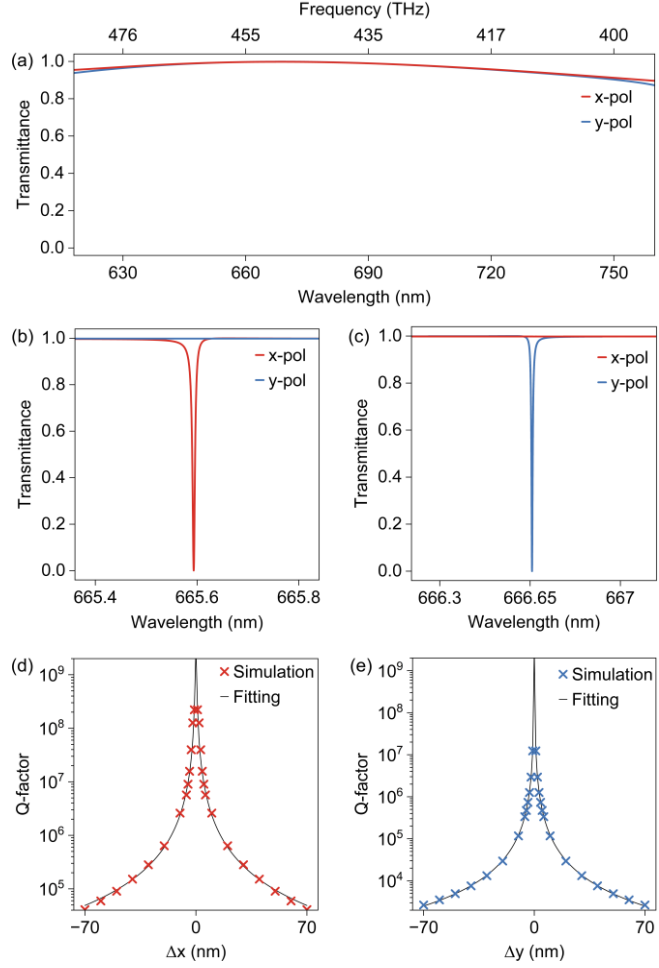

**Fig. 3.** The transmittance spectra under normal incident linearly polarized light for (a) $\Delta x$=0 nm and $\Delta y$=0 nm, (b) $\Delta x$=45 nm and $\Delta y$=0 nm, (c) $\Delta x$=0 nm and $\Delta y$=10 nm. The Q-factors of the quasi-BIC mode as a function of (d) $\Delta x$ and (e) $\Delta y$.

To investigate the optical response of the structure, the transmission enabled by the quasi-BIC resonance is calculated. Under normal incident linearly polarized light, the structure with $\Delta x$=0 nm and $\Delta y$=0 nm is perfectly transparent in the range of the concerned wavelengths. As shown in Fig. 3(a), the transmittance spectra show a wideband totally transmission Fabry-Perot background due to the wavevector mismatch between the guided mode and external incident light. Following the selection rules, we can manipulate the quasi-BIC radiation to different polarization directions by moving one of the rectangular bars along the $x$ or $y$ directions. Figs. 3(b) and 3(c) are the transmittance spectra of the structure for $\Delta x$=45 nm and $\Delta y$=10 nm under the normal incident linearly polarized light, respectively. When one of the rectangular bars moves along the $x$ direction, the $x$-polarized light leads to a leaky resonance in the transmissive Fabry-Perot background, as shown in Fig. 3(b). In contrast, when the rectangular bar only moves along the $y$ direction, the $y$-polarized light results in a leaky resonance, as shown in Fig. 3(c). Furthermore, Figs. 3(d) and 3(e) illustrate the Q-factors as a function of $\Delta x$ and $\Delta y$, respectively. The Q-factors of the quasi-BIC resonance experience a dramatic increase as $\Delta x$ or $\Delta y$ decreases and diverge to infinity near $\Delta x$=0 nm or $\Delta y$=0 nm. According to the signature of quasi-BICs, its Q-factor approximately follows an inversely quadratic law of asymmetric perturbations[13]. This relation is verified by simulation results in Fig.3(d) and 3(e), the scale factors of quasi-BIC are retrieved to be $2.4*10^8$ and $1.2*10^7$ for $\Delta x$ and $\Delta y$, respectively. Here, it can be noticed that the Q-factors of the quasi-BIC resonance exhibit different sensitivities to $\Delta x$ and $\Delta y$, and the scale factor is calculated to be $g = \sqrt{\frac{2.4*10^8}{1.2*10^7}} = 4.48$. Consequently, when $\Delta x = g\Delta y$, the $|m_x| = |m_y|$ will be achieved in the structure. Importantly, there is a π/2 phase difference between $m_x$ and $m_y$ because the quasi-BIC is aligned to a transmissive Fabry-Perot background and the structure lacks z-axis mirror symmetry[34].

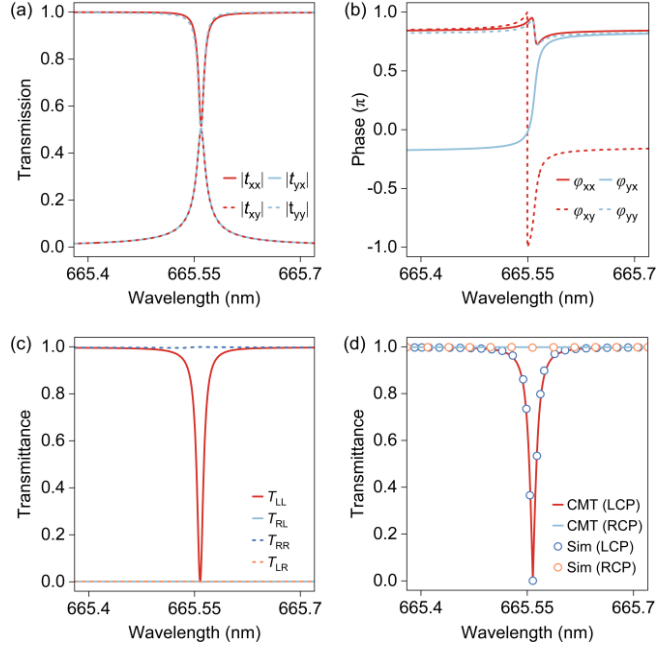

**Fig. 4.** (a) The transmission and (b) phase spectra under normal incident linearly polarized light for $\Delta x$=45 nm and $\Delta y$=10 nm. The corresponding transmittance spectra of the (c) all components of the circularly polarized light, (d) RCP and LCP light

In order to validate the above analysis, the amplitudes and phases of the structure under linearly polarized incidence are calculated, as shown in Fig. 4(a) and Fig. 4(b), respectively. Here, we choose $\Delta x$=45 nm and $\Delta y$=10 nm according to the different sensitivities of Q-factors to $\Delta x$ and $\Delta y$. It can be found that the amplitudes of the co-polarization and cross-polarization transmission coefficients are identical $|t_{xx}| = |t_{yx}| = |t_{yy}| = |t_{xy}| = 0.5$ at the quasi-BIC resonance wavelength. Meanwhile, the phase difference between $t_{xx}$ and $t_{yy}$ is $\varphi_{xx} = \varphi_{yy}$, and the phase difference between $t_{yy}$ and $t_{yx}$ is $\varphi_{yy} - \varphi_{yx} = \pi/2$.

As a consequence, the quasi-BIC is transformed into chiral quasi-BIC. As shown in Fig. 4(c), the RCP light decoupled from the chiral quasi-BIC in the far field, while the LCP light remains strongly coupled and there is no polarization conversion within this interaction. The nearly perfect spin-selective transmission is achieved at the resonant wavelength of λ=665.55 nm, the CD reaches near unity 0.9988. As shown in Fig. 4(d), the simulated transmittance curve can be well fitted by Eq. (12).

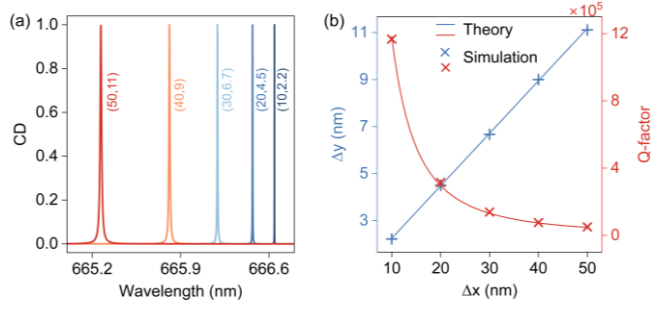

**Fig. 5.** (a) The CD of the chiral quasi-BIC as functions of the cooperative change of perturbations $\Delta x$ and $\Delta y$. (b) The simulated and theoretically predicted relation between $\Delta x$ and $\Delta y$ for maximizing CD and $Q$-factor of the chiral quasi-BIC resonance.

In some of previous works, there exists an inherent limit between the $Q$-factors and CD of the chiral quasi-BIC, i.e., an increase in the asymmetric parameters on the one hand leads to an increase in the CD but on the other hand an exponential increase in the radiation loss[57]. However, this limitation can be broken in our design. According to the above discussion, the intrinsic chirality of the quasi-BIC can be maintained, as long as the two perturbations $\Delta x$ and $\Delta y$ are changed cooperatively. The $Q$-factor of the chiral quasi-BIC resonance in the designed planar chiral structure can be, in principle, continuously boosted while the amplitude of the CD remains near unity, as shown in Fig. 5(a). What's more, the inherent linkage between $\Delta x$ and $\Delta y$ for the maximal intrinsic chirality can be theoretically predicted by $|m_x| = |m_y|$, following a linear relationship of $\Delta x = g\Delta y$ (fitted by blue line), which is confirmed by simulation results in Fig. 5(b). Here, the scale factor is related to the mode profile and could take different values for different chiral quasi-BICs. The $Q$-factor of quasi-BICs follows the $Q \sim 1/(\Delta x^2 + g^2\Delta y^2)$ (fitted by red line) roughly scales with the inversely quadratic square of all the perturbations. In the present design, the change in the $Q$-factor exceeds two orders of magnitude.

## Conclusions

In conclusion, we propose a straightforward method to achieve the extreme intrinsic chirality in planar structures through manipulating the quasi-BIC *via* in-plane perturbation, and we demonstrate it in a two-dimensional waveguide grating composed of a dimeric grating and a waveguide layer. The CMT is initially employed to derive the conditions necessary for achieving maximal intrinsic chirality in lossless structures. Benefiting from the precisely controlling of the introduced in-plane perturbation to the coupling of quasi-BIC with linearly polarized waves, the high-$Q$ guided mode resonance corresponding to the quasi-BIC is excited within the transparent spectral range of the structure, and coupled with *x*- and *y*-polarized waves with the same intensity but a phase difference of $\pi/2$, ultimately resulting in the attainment of the extreme intrinsic chirality. The simulation results demonstrate the quasi-BIC exhibits strong interaction with LCP light while decoupling from RCP light, with a CD value of 0.9988. More interestingly, by adjusting the magnitude of the in-plane asymmetry, the $Q$-factors of the chiral quasi-BIC are independently manipulated exceeds two orders of magnitude while the CD is maintained nearly unity. Our results provide a simple but powerful paradigm of achieving strong chiral light-matter interactions on an easy-fabricated platform, which could offer the possibility of designing high-performance spin-selective optical devices such as chiral emitters, sensors, photonic circuits, etc.

**Funding**

This work was supported by the National Natural Science Foundation of China (Grants Nos. 12064025, 11604136, 12304420, 12264028, 12364045, 12204552), the Natural Science Foundation of Jiangxi Province (Grants Nos. 20212ACB202006, 20232BAB201040, 20232BAB211025), the Young Elite Scientists Sponsorship Program by JXAST (Grant No.2023QT11), and Innovation Fund for Graduate Students of Jiangxi Province (Grant No. YC2022-B019).

**Notes**

The authors declare no competing financial interest